\documentclass{PoS}
\usepackage{amsmath,pstricks}

\title{On the resummation of non-global logarithms at finite $\mathbf{N_c}$}

\ShortTitle{On the resummation of NGLs at finite $N_c$}

\author{\speaker{Yazid DELENDA}\\
        D\'{e}partement des Sciences de la Mati\`{e}re, Facult\'{e} des Sciences\\
        Universit\'{e} Hadj Lakhdar - Batna, Algeria\\
        E-mail: \email{yazid.delenda@gmail.com}}

\author{Kamel KHELIFA-KERFA\\
        D\'{e}partement de Physique, Facult\'{e} des Sciences\\
        Universit\'{e} Hassiba Benbouali de Chlef - Chlef, Algeria\\
        E-mail: \email{kamel.kkhelifa@gmail.com}}

\abstract{We present a calculation of non-global logs at finite $\mathrm{N_c}$ for the hemisphere mass distribution in $e^+e^-\to 2$ jets at single log accuracy up to fifth order in the strong coupling constant. Our results suggest a possible all-orders resummation of these large logs into an exponential. Comparing our results to those at large $\mathrm{N_c}$, recently reported in literature, we find an agreement. We additionally compare our findings with the numerical all-orders resummation at large $\mathrm{N_c}$ and discuss the significance of neglected finite-$\mathrm{N_c}$ corrections on the said distribution.}

\FullConference{XXIII International Workshop on Deep-Inelastic Scattering and Related Subjects\\
        		April 27 - May 1, 2015\\
		        Dallas, Texas}

\begin{document}

\section{Introduction}

With the restart of the LHC at an unprecedented 13 TeV center-of-mass energy, the search for beyond-standard-model particles continues. For a successful achievement of the goals of this second run, precision is vital from both theory calculations and experimental measurements. Due to the inevitable hadronic environment involved at the LHC, where for example most Higgs studies are overwhelmed by QCD background, analytic estimates of QCD observable cross-sections will continue to play a central role leading either to the systematic improvement/tuning of Monte Carlo event generators, or to the development of better methods of background elimination.

The resummation of large logs, resulting from the real/virtual mis-cancellation of soft and/or collinear singularities in the matrix element, is perhaps the most challenging QCD perturbative aspect when one attempts to make an estimate of the cross-section of a given observable $V$. For observables that are sensitive to emissions in the entire angular phase space, and which are termed ``global'' observables, the resummed distribution maybe cast into the general form:
\begin{equation}\label{eq:resum}
\sigma(V) \propto \exp\left(L g_1(\alpha_s L)+g_2(\alpha_s L)+\alpha_s g_3(\alpha_sL)+\alpha_s^2 g_4(\alpha_sL)+\cdots\right),
\end{equation}
where $L$ is the large log of the observable $V$. The functions $g_1$, $g_2$, $\dots$, respectively resum leading logs (LL), next-to-leading logs (NLL), $\dots$.

While the development of the resummation programme for global observables has seen substantial progress in recent years, achieving up to NNNLL accuracy (i.e. up to $g_4$ in eq. \eqref{eq:resum}), e.g. for the $C$-parameter distribution \cite{Hoang:2014wka}, and even semi-automatic resummation to NLL \cite{Banfi:2004yd} and NNLL \cite{Banfi:2014sua}, progress in the resummation of ``non-global'' observables \cite{Dasgupta:2001sh,Dasgupta:2002bw} has been very limited. Non-global observables are those which are sensitive to emissions in restricted regions of the angular phase space, and as a result their distributions suffer from non-global logs (NGLs). There are several important observables that are non-global and that are widely used in studies relevant to new physics searches, such as jet mass and single hemisphere observables.

There are two main reasons that have long jeopardised progress in the resummation of NGLs. Firstly, the treatment of cascade gluon branching is extremely cumbersome within perturbation theory. Secondly, the phase-space integrations that one has to perform are prohibitive at higher orders due to non-iterative geometry involved in the calculation. A practical solution to the above hindrances is resorting to the large-$\mathrm{N_c}$ approximation, with $\mathrm{N_c}$ being the number of quark colours, which amounts to discarding non-planar Feynman diagrams \cite{'tHooft:1973jz} in the calculation of amplitudes of  soft gluon emissions, leading to considerable simplifications. Specifically this approximation is equivalent to the leading-order expansion of the colour factor $\mathrm{C_F} = \mathrm{N_c}/2-1/2\mathrm{N_c} \approx \mathrm{N_c} /2$. Furthermore, a convenient approach to deal with the multi-dimensional integrations is restoring to numerical Monte Carlo methods. The numerical resummation of NGLs at large $\mathrm{N_c}$ was first performed in the pioneering work of Dasgupta and Salam \cite{Dasgupta:2001sh}.

There has recently been an increasing interest in literature in the calculation of NGLs both at fixed order and to all orders. In ref. \cite{Rubin:2010fc}, Rubin numerically evaluated NGLs at large $\mathrm{N_c}$ for both filtered Higgs jet mass and interjet energy flow up to sixth order in $\alpha_s$. Hatta and Ueda \cite{Hatta:2013iba} performed a numerical resummation of NGLs based on the Weigert equation \cite{Weigert:2003mm}, accounting for NGLs at finite $\mathrm{N_c}$ to all orders. A similar equation that was developed by Banfi, Marchesini and Smye, the BMS equation \cite{Banfi:2002hw}, and whose solution accounts for the all-orders resummation of NGLs at large $\mathrm{N_c}$, was the subject of study by Schwartz and Zhu in ref. \cite{Schwartz:2014wha}, where the analytic solution up to fifth order was achieved.

The aim of this work is to address the question of how the accuracy of resummation of NGLs is affected by the large-$\mathrm{N_c}$ approximation. For this, and other reasons, we perform the analytic calculation of NGLs at finite $\mathrm{N_c}$ up to fifth order. In the next section we define the observable that we use to illustrate the calculation of NGLs, namely the single-hemisphere mass distribution in $e^+e^-\to q\bar{q}$. In section 3 we show the results for the NGLs up to fifth order, which are then used to make an anstaz for the all-orders (partial) resummation of NGLs into an exponential form. We also compare, in the same section, our findings to those reported at large $\mathrm{N_c}$  by Schwartz and Zhu \cite{Schwartz:2014wha}. In section 4 we perform a comparison with the numerical results obtained by Dasgupta and Salam \cite{Dasgupta:2001sh} and discuss the implications of our results. Finally we summarise and give future directions of this work in section 5.

\section{Observable and kinematics}

We are interested in the calculation of NGLs at finite $\mathrm{N_c}$ up to fifth order. For illustrative purposes we choose to study the simple process $e^+e^- \to q\bar{q}$ accompanied by the emission of soft energy-ordered gluons $k_i$, as depicted in figure \ref{fig:setup}.
\input{Figs/fig1.tex}
The quark and anti-quark directions determine two back-to-back hemispheres $\mathcal{H}_L$ and $\mathcal{H}_R$. We consider for measurement the right hemisphere $\mathcal{H}_R$ and calculate its normalised invariant mass $\rho$ defined by:
\begin{align}
\rho =& \left(p_q + \sum_{i\in \mathcal{H}_R} k_i\right)^2/Q^2 \approx 2 \sum_{i\in \mathcal{H}_R} k_i \cdot p_q/Q^2\,,
\label{eq:def}
\end{align}
where $Q$ is the center-of-mass energy and $p_q$ and $k_i$ are the four-momenta of the quark and gluons, respectively.

The integrated hemisphere-mass distribution, normalised to the Born cross-section $\sigma_0$, is:
\begin{subequations}\label{eq:distribution}
\begin{align}
\sigma(\rho) &= \int_0^\rho \frac{1}{\sigma_0} \frac{\mathrm{d}\sigma}{\mathrm{d}\rho'}\mathrm{d}\rho' = 1+\sigma_1(\rho) + \sigma_2(\rho)+\cdots\,,\\
\sigma_m &= \sum_X \int \mathrm{d}\phi_m \hat{\mathcal{U}}_m \mathcal{W}_{12\cdots m}^X\,,
\end{align}
\end{subequations}
where $\mathcal{W}_{12\cdots m}^X = \mathcal{W}^X(k_1,k_2,\dots, k_m)$ represents the eikonal squared amplitude, normalised to the Born squared amplitude, for the emission of $m$ energy-ordered gluons and $\mathrm{d}\phi_m$ is the corresponding phase space. The sum over $X$ accounts for all possible real/virtual configurations of the radiated gluons. The \emph{measurement operator} $\hat{\mathcal{U}}_m$ plays the role of an event selector, i.e. it forbids real emissions into $\mathcal{H}_R$ which contribute more than $\rho$ to the hemisphere mass.

In order to be able to compute the cross-section \eqref{eq:distribution}, there are several issues that need to be addressed. First, one must evaluate the gluon-emission squared amplitudes $\mathcal{W}_{12\cdots m}^X$, including all the possible real-virtual configurations $X$ at each order in the perturbation series (our aim is up to fifth order). The calculation of such amplitudes is non-trivial due to the complexity of the colour algebra involved as well as the factorially growing number of Feynman diagrams that one has to account for at each order. Further details about the computation of these eikonal amplitudes, which involves using the \texttt{Mathematica} package \texttt{ColorMath} \cite{Sjodahl:2012nk}, are to be found in our work in refs. \cite{Khelifa:2015,Delenda:2015}.

The second task is to apply the measurement operator according to the various real-virtual gluon configurations in order to extract the appropriate phase-space region of integration for each gluon. Doing so the final task is to perform the relevant multi-dimensional integrations in eq. \eqref{eq:distribution}. At fourth order, for instance, the integral is seven-dimensional and was performed semi-analytically.

\section{Non-global logs up to fifth order at finite $\mathbf{N_c}$}

We can write the integrated hemisphere mass distribution \eqref{eq:distribution} as:
\begin{subequations}\label{eq:primary}
\begin{align}
\sigma(\rho) & = \sigma^\mathrm{S}(\rho) \times \sigma^\mathrm{NG}(\rho)\,, \\
\sigma^\mathrm{S}(\rho) & =  \exp\left(-\mathrm{C_F}\bar{\alpha}_s L^2\right),
\end{align}
\end{subequations}
where $L=\ln(1/\rho)$ and $\bar{\alpha}_s = \alpha_s/\pi$. We have factorised the distribution into the product of a Sudakov form factor $\sigma^\mathrm{S}$, that resums double logs originating from soft-collinear primary emissions, times a non-global factor $\sigma^\mathrm{NG}$, that resums single logs originating from soft wide-angle secondary correlated emissions.

We express the non-global factor as a series in the coupling starting from second order, where NGLs first appear, up to fifth order as follows:
\begin{align}
\sigma^\mathrm{NG}(\rho) =& 1 - \frac{\bar{L}^2}{2!}\mathrm{C_F} \mathrm{C_A} \zeta_2 + \frac{\bar{L}^3}{3!}\mathrm{C_F} \mathrm{C_A^2} \zeta_3 -
\frac{\bar{L}^4}{4!}\left(\frac{25}{8}\mathrm{C_F} \mathrm{C_A^3}\zeta_4 - \frac{13}{5} \mathrm{C_F^2} \mathrm{C_A^2}\zeta_2^2\right)-  \notag \\
  &- \frac{\bar{L}^5}{2!3!}\mathrm{C_F^2} \mathrm{C_A^3}\zeta_2 \zeta_3+ \frac{\bar{L}^5}{5!}\mathrm{C_F} \mathrm{C_A^4}\zeta_5 \left[\alpha + \beta \left(\frac{\mathrm{C_F}}{\mathrm{C_A}}-\frac{1}{2}\right)\right]+\mathcal{O}(\alpha_s^6)\,, \label{eq:NGLsseries}
\end{align}
with $\bar{L} = \bar{\alpha}_sL$, $\mathrm{C_A}=\mathrm{N_c}$, and $\alpha$ and $\beta$ are constants that are yet to be determined.

To compare our result \eqref{eq:NGLsseries} with that obtained in ref. \cite{Schwartz:2014wha} at large $\mathrm{N_c}$, by means of analytic solution to the BMS equation, we simply make the substitution $\mathrm{C_F}\to \mathrm{N_c}/2$, leading to:
\begin{align}
\sigma^\mathrm{NG}(\rho) =& 1 - \frac{\pi^2}{24}(\mathrm{N_c}\bar{L})^2 +\frac{\zeta_3}{12} (\mathrm{N_c}\bar{L})^3+
\frac{\pi^4}{34\,560}(\mathrm{N_c}\bar{L})^4+\left(-\frac{\pi^2\zeta_3}{288}+\alpha \frac{\zeta_5}{240}\right)(\mathrm{N_c}\bar{L})^5+\mathcal{O}(\alpha_s^6)\,,\label{eq:large-nc}
\end{align}
which exactly agrees with the result arrived at in ref. \cite{Schwartz:2014wha} up to fourth order. Furthermore, at fifth order we can extract the value of the undetermined constant $\alpha$ by comparison with the result in ref. \cite{Schwartz:2014wha} and we obtain $\alpha =17/2+\zeta_2\zeta_3/\zeta_5$. We can also further make an anstaz for the constant $\beta$ based on the pattern of zeta functions observed at previous orders: $\beta = 2\zeta_2 \zeta_3/\zeta_5$.

In order to check the impact of finite-$\mathrm{N_c}$ corrections (relative to large-$\mathrm{N_c}$ result) on the distribution, we compare the result at large $\mathrm{N_c}$ (eq. \eqref{eq:large-nc}) to that at finite $\mathrm{N_c}$ (eq. \eqref{eq:NGLsseries}). We find that at fourth order the size of finite-$\mathrm{N_c}$ result constitutes merely an $\mathcal{O}(1.5\%)$ correction to the large-$\mathrm{N_c}$ result. This observation is in accordance with that made in ref. \cite{Hatta:2013iba} by means of numerical evaluation of NGLs at finite $\mathrm{N_c}$ through a solution to the Weigert equation.

We note that the series of NGLs in eq. \eqref{eq:NGLsseries} adheres a pattern of an expansion of an exponential function:
\begin{align}
\sigma^{\mathrm{NG}}(\rho) =& \exp\left(-\frac{\bar{L}^2}{2!}\mathrm{C_FC_A}\zeta_2+\frac{\bar{L}^3}{3!}\mathrm{C_FC_A^2}\zeta_3-\frac{\bar{L}^4}{4!}\mathrm{C_FC_A^3}\zeta_4
\left[\frac{29}{8}+\left(\frac{\mathrm{C_F}}{\mathrm{C_A}}-\frac{1}{2}\right)\right]+\right. \notag\\
   &\quad\quad\left.+ \frac{\bar{L}^5}{5!}\mathrm{C_FC_A^4}\zeta_5
\left[\alpha+\beta\left(\frac{\mathrm{C_F}}{\mathrm{C_A}}-\frac{1}{2}\right)\right]+\mathcal{O}(\alpha_s^6)\right)\notag\\
=& \exp\left(-\frac{\bar{L}^2}{2!}\mathrm{C_FC_A}\zeta_2+\frac{\bar{L}^3}{3!}\mathrm{C_FC_A^2}\zeta_3-\frac{\bar{L}^4}{4!}
\left[  \frac{25}{8}\mathrm{C_FC_A^3}\zeta_4+\frac{2}{5}\mathrm{C_F^2C_A^2}\zeta_2^2\right]+\right. \notag\\
   &\quad\quad\left.+ \frac{\bar{L}^5}{5!}
\left[\frac{17}{2}\mathrm{C_FC_A^4}\zeta_5+2\mathrm{C_F^2C_A^3}\zeta_2\zeta_3\right]+\mathcal{O}(\alpha_s^6)\right), \label{eq:exponential}
\end{align}
where we substituted the values of $\alpha$ and $\beta$ based on the observations discussed above. The first form of the above exponential explicitly displays the finite-$\mathrm{N_c}$ corrections, whilst the second form focuses on disclosing the pattern of the Zeta functions.

To determine the convergence of the series in the exponent of eq. \eqref{eq:exponential}, we plot in figure \ref{fig2}
\begin{figure}[ht]
\begin{center}
\scalebox{1}{
\begin{pspicture}(0,-3.060625)(8.658907,3.060625)
\definecolor{color1}{rgb}{0.3686274509803922,0.5058823529411764,0.7098039215686275}
\definecolor{color147}{rgb}{1.0,0.4,0.0}
\definecolor{color1687}{rgb}{0.4,0.4,1.0}
\definecolor{color148}{rgb}{0.4,0.6,0.0}
\psline[linewidth=0.038399998,linecolor=color1687](0.60546875,-1.0865625)(0.7728906,-1.0865625)(0.7824219,-1.0864844)(0.868125,-1.0864844)(0.87757814,-1.0864062)(0.9202344,-1.0864062)(0.9253125,-1.0863281)(0.96101564,-1.0863281)(0.9660938,-1.08625)(0.97625,-1.08625)(0.9966406,-1.0861719)(1.0157031,-1.0861719)(1.0346875,-1.0860938)(1.0442188,-1.0860938)(1.05375,-1.0860156)(1.0727344,-1.0859375)(1.0830469,-1.0859375)(1.0933594,-1.0858594)(1.1139843,-1.0857812)(1.1242969,-1.0857812)(1.1346093,-1.0857031)(1.1552343,-1.085625)(1.1603125,-1.085625)(1.1653125,-1.0855469)(1.1754688,-1.0855469)(1.1957031,-1.0853906)(1.2361718,-1.0851562)(1.2409375,-1.0851562)(1.245625,-1.0850781)(1.2550781,-1.085)(1.2739062,-1.0849218)(1.3117187,-1.0846094)(1.3167969,-1.0846094)(1.3219532,-1.0845313)(1.3526562,-1.0842968)(1.3935938,-1.0839063)(1.3960156,-1.0839063)(1.3983594,-1.0838281)(1.403125,-1.0838281)(1.4127344,-1.0836719)(1.4317969,-1.0835156)(1.4700781,-1.0830469)(1.545,-1.0821093)(1.6263281,-1.0809375)(1.7021875,-1.0796875)(1.7844532,-1.078125)(1.8651563,-1.0764062)(1.9404688,-1.0746094)(2.0221093,-1.0724219)(2.0983593,-1.0701562)(2.1809375,-1.0675)(2.2620313,-1.0646094)(2.3377345,-1.0617187)(2.4197657,-1.0582813)(2.496328,-1.0548438)(2.5714064,-1.05125)(2.6528907,-1.0470313)(2.7289062,-1.0428907)(2.81125,-1.0380468)(2.8921094,-1.0329688)(2.9675782,-1.0278906)(3.049375,-1.0221875)(3.1257813,-1.0164844)(3.200625,-1.0105469)(3.281875,-1.0038282)(3.3576562,-0.99726564)(3.4397657,-0.9896875)(3.5164845,-0.9822656)(3.5917187,-0.9746875)(3.6733594,-0.96601564)(3.749453,-0.9575)(3.831953,-0.94789064)(3.9129689,-0.93796873)(3.9885938,-0.92828125)(4.070547,-0.91734374)(4.1470313,-0.90664065)(4.222031,-0.89570314)(4.3033595,-0.88328123)(4.379297,-0.8711719)(4.461641,-0.8575)(4.542422,-0.84351563)(4.6177344,-0.82984376)(4.6994534,-0.814375)(4.775781,-0.799375)(4.8584375,-0.78234375)(4.9395313,-0.7649219)(5.0152345,-0.7480469)(5.097344,-0.72882813)(5.1739845,-0.7102344)(5.2491407,-0.6911719)(5.330625,-0.6696875)(5.4067187,-0.64867187)(5.4891405,-0.625)(5.570078,-0.60078126)(5.645547,-0.5771875)(5.7274218,-0.55046874)(5.8038282,-0.52453125)(5.87875,-0.49796876)(5.96,-0.46796876)(6.0358596,-0.4388281)(6.1180468,-0.40578127)(6.194844,-0.37359375)(6.270078,-0.34070313)(6.351719,-0.3035156)(6.427969,-0.26734376)(6.5105467,-0.22632812)(6.5915623,-0.18421876)(6.6671877,-0.14328127)(6.749219,-0.09695312)(6.8257813,-0.05171876)(6.900781,-0.00554688)(6.9821873,0.04679688)(7.058203,0.0978125)(7.140547,0.15554687)(7.2214065,0.21492188)(7.296797,0.27273437)(7.378594,0.33820313)(7.4549217,0.40210938)(7.5297656,0.4674219)(7.6109376,0.54148436)(7.686719,0.61375)(7.768828,0.69570315)(7.8455467,0.77570313)(7.920703,0.85765624)(8.002266,0.95054686)(8.07836,1.0413281)(8.160859,1.1442187)(8.2417965,1.2501563)(8.317344,1.3535937)(8.317422,1.35375)(8.317578,1.3538281)(8.318672,1.3554688)(8.32,1.3572656)(8.322657,1.3609375)(8.33836,1.3832031)(8.359375,1.413125)(8.359453,1.4132812)(8.35961,1.4133594)(8.360078,1.4140625)(8.360703,1.4150782)(8.364688,1.4207032)(8.369922,1.4282812)(8.38039,1.4435157)(8.380625,1.44375)(8.381719,1.4453906)(8.383047,1.4472656)(8.385703,1.4510938)(8.390938,1.45875)(8.391016,1.4589063)(8.391094,1.4589844)(8.39125,1.4592968)(8.391562,1.4597657)(8.392265,1.4607031)(8.393594,1.4626563)(8.396172,1.4664844)(8.39625,1.4665625)(8.396328,1.4667188)(8.396484,1.4669532)(8.396875,1.4674219)(8.3975,1.4683594)(8.3988285,1.4703125)(8.398907,1.4704688)(8.398984,1.4705468)(8.399453,1.47125)(8.400156,1.4722656)(8.400234,1.4723437)(8.400312,1.4725)(8.400782,1.4732031)(8.40086,1.4733593)(8.400937,1.4734375)(8.4010935,1.4736719)(8.401172,1.4738281)(8.40125,1.4739063)(8.401328,1.4740624)(8.401406,1.4741406)
\psline[linewidth=0.038399998,linecolor=color147](0.60546875,-1.0865625)(0.7728906,-1.0865625)(0.7824219,-1.0864844)(0.8634375,-1.0864844)(0.868125,-1.0864062)(0.91765624,-1.0864062)(0.9202344,-1.0863281)(0.9565625,-1.0863281)(0.9571875,-1.08625)(0.97625,-1.08625)(0.9966406,-1.0861719)(1.0061718,-1.0861719)(1.0157031,-1.0860938)(1.0394531,-1.0860938)(1.0442188,-1.0860156)(1.05375,-1.0860156)(1.0727344,-1.0859375)(1.0778906,-1.0859375)(1.0830469,-1.0858594)(1.0933594,-1.0858594)(1.1139843,-1.0857812)(1.1153125,-1.0857812)(1.1165625,-1.0857031)(1.1242969,-1.0857031)(1.1346093,-1.085625)(1.1552343,-1.0855469)(1.1603125,-1.0855469)(1.1653125,-1.0854688)(1.1754688,-1.0853906)(1.1957031,-1.0853125)(1.1982031,-1.0853125)(1.2007812,-1.0852344)(1.2057812,-1.0852344)(1.2159375,-1.0851562)(1.2361718,-1.085)(1.2409375,-1.085)(1.245625,-1.0849218)(1.2550781,-1.0848438)(1.2739062,-1.0847657)(1.3117187,-1.084375)(1.3167969,-1.084375)(1.3219532,-1.0842968)(1.3321875,-1.0842187)(1.3526562,-1.0839844)(1.3935938,-1.0835937)(1.3942188,-1.0835937)(1.3947656,-1.0835156)(1.3983594,-1.0835156)(1.403125,-1.0834374)(1.4127344,-1.0833594)(1.4317969,-1.083125)(1.4700781,-1.0825781)(1.545,-1.0814844)(1.6263281,-1.0800781)(1.7021875,-1.0785156)(1.7844532,-1.0765625)(1.8651563,-1.0742968)(1.9404688,-1.0720313)(2.0221093,-1.0691407)(2.0983593,-1.0661719)(2.1809375,-1.0626563)(2.2620313,-1.05875)(2.3377345,-1.0547656)(2.4197657,-1.05)(2.496328,-1.0451562)(2.5714064,-1.04)(2.6528907,-1.0339844)(2.7289062,-1.0278906)(2.81125,-1.0207031)(2.8921094,-1.0132031)(2.9675782,-1.0057031)(3.049375,-0.9969531)(3.1257813,-0.9882031)(3.200625,-0.97914064)(3.281875,-0.96867186)(3.3576562,-0.9582813)(3.4397657,-0.94640625)(3.5164845,-0.93460935)(3.5917187,-0.9224219)(3.6733594,-0.9084375)(3.749453,-0.8947656)(3.831953,-0.8790625)(3.9129689,-0.8628125)(3.9885938,-0.8469531)(4.070547,-0.82890624)(4.1470313,-0.8111719)(4.222031,-0.7930469)(4.3033595,-0.7725)(4.379297,-0.75242186)(4.461641,-0.7296875)(4.542422,-0.7063281)(4.6177344,-0.68359375)(4.6994534,-0.6578906)(4.775781,-0.63289064)(4.8584375,-0.6047656)(4.9395313,-0.5759375)(5.0152345,-0.5478906)(5.097344,-0.51640624)(5.1739845,-0.48578125)(5.2491407,-0.4546875)(5.330625,-0.4196875)(5.4067187,-0.38578126)(5.4891405,-0.3477344)(5.570078,-0.30890626)(5.645547,-0.27140626)(5.7274218,-0.22929688)(5.8038282,-0.18859376)(5.87875,-0.14734375)(5.96,-0.10109376)(6.0358596,-0.05640626)(6.1180468,-0.00632812)(6.194844,0.04203124)(6.270078,0.0909375)(6.351719,0.14578123)(6.427969,0.19859374)(6.5105467,0.2578125)(6.5915623,0.3178125)(6.6671877,0.375625)(6.749219,0.4403125)(6.8257813,0.50257814)(6.900781,0.5655469)(6.9821873,0.6360156)(7.058203,0.70375)(7.140547,0.77953124)(7.2214065,0.85625)(7.296797,0.93007815)(7.378594,1.0125)(7.4549217,1.091875)(7.5297656,1.1719531)(7.6109376,1.2614062)(7.686719,1.3474219)(7.768828,1.4435157)(7.8455467,1.5359375)(7.920703,1.6291406)(8.002266,1.7332813)(8.07836,1.8334374)(8.160859,1.9451562)(8.2417965,2.0582812)(8.317344,2.166953)(8.317422,2.1671095)(8.317578,2.1671875)(8.318047,2.1678905)(8.318672,2.1689062)(8.32,2.1707811)(8.322657,2.1746094)(8.32789,2.1823437)(8.33836,2.1977344)(8.359375,2.22875)(8.359453,2.2289062)(8.35961,2.2289844)(8.360078,2.2296875)(8.360703,2.230703)(8.364688,2.2365625)(8.38039,2.26)(8.380468,2.2601562)(8.380625,2.2602344)(8.380781,2.2604687)(8.381094,2.2610157)(8.381719,2.261953)(8.383047,2.2639062)(8.385703,2.2678907)(8.390938,2.2757032)(8.391016,2.2758594)(8.391094,2.2759376)(8.39125,2.27625)(8.391562,2.2767189)(8.392265,2.2777343)(8.393594,2.2796874)(8.396172,2.2835937)(8.39625,2.28375)(8.396328,2.283828)(8.396484,2.2841406)(8.396875,2.2846093)(8.3975,2.2855468)(8.3988285,2.287578)(8.398907,2.2876563)(8.398984,2.2878125)(8.399453,2.2885156)(8.400156,2.2895312)(8.400234,2.2896874)(8.400312,2.2897656)(8.400469,2.29)(8.400782,2.290547)(8.40086,2.290625)(8.400937,2.2907813)(8.4010935,2.2910156)(8.401172,2.2911718)(8.40125,2.29125)(8.401328,2.2914062)(8.401406,2.2914844)
\psline[linewidth=0.038399998,linecolor=color148](0.60546875,-1.0865625)(0.7728906,-1.0865625)(0.7824219,-1.0864844)(0.868125,-1.0864844)(0.87757814,-1.0864062)(0.9202344,-1.0864062)(0.9253125,-1.0863281)(0.96101564,-1.0863281)(0.9660938,-1.08625)(0.97625,-1.08625)(0.9966406,-1.0861719)(1.0157031,-1.0861719)(1.0346875,-1.0860938)(1.0442188,-1.0860938)(1.05375,-1.0860156)(1.0727344,-1.0859375)(1.0830469,-1.0859375)(1.0933594,-1.0858594)(1.1139843,-1.0857812)(1.1242969,-1.0857812)(1.1346093,-1.0857031)(1.1552343,-1.085625)(1.1603125,-1.085625)(1.1653125,-1.0855469)(1.1754688,-1.0855469)(1.1957031,-1.0853906)(1.2361718,-1.0851562)(1.2409375,-1.0851562)(1.245625,-1.0850781)(1.2550781,-1.0850781)(1.2739062,-1.0849218)(1.3117187,-1.0846094)(1.3167969,-1.0846094)(1.3219532,-1.0845313)(1.3526562,-1.0842968)(1.3935938,-1.0839063)(1.3983594,-1.0839063)(1.403125,-1.0838281)(1.4127344,-1.08375)(1.4317969,-1.0835156)(1.4700781,-1.083125)(1.545,-1.0821875)(1.6263281,-1.0810938)(1.7021875,-1.0798438)(1.7844532,-1.0783594)(1.8651563,-1.0767188)(1.9404688,-1.075)(2.0221093,-1.0729687)(2.0983593,-1.0709375)(2.1809375,-1.0684375)(2.2620313,-1.0658594)(2.3377345,-1.0632813)(2.4197657,-1.0602344)(2.496328,-1.0572656)(2.5714064,-1.0541406)(2.6528907,-1.050625)(2.7289062,-1.0471874)(2.81125,-1.0432031)(2.8921094,-1.0392188)(2.9675782,-1.0353125)(3.049375,-1.0308594)(3.1257813,-1.0266407)(3.200625,-1.0223438)(3.281875,-1.0176562)(3.3576562,-1.0130469)(3.4397657,-1.0080469)(3.5164845,-1.0032812)(3.5917187,-0.9985156)(3.6733594,-0.9933594)(3.749453,-0.9884375)(3.831953,-0.983125)(3.9129689,-0.9778906)(3.9885938,-0.97296876)(4.070547,-0.9677344)(4.1470313,-0.9628906)(4.222031,-0.95820314)(4.3033595,-0.9532813)(4.379297,-0.94875)(4.461641,-0.9439844)(4.542422,-0.9394531)(4.6177344,-0.93546873)(4.6994534,-0.9313281)(4.775781,-0.9277344)(4.8584375,-0.92414063)(4.9395313,-0.9209375)(5.0152345,-0.91828126)(5.0165625,-0.91828126)(5.0178127,-0.9182031)(5.025547,-0.91796875)(5.0357814,-0.91765624)(5.0563283,-0.91695315)(5.097344,-0.91578126)(5.0985155,-0.91578126)(5.099766,-0.9157031)(5.1021094,-0.915625)(5.106953,-0.9155469)(5.116484,-0.9152344)(5.135625,-0.9147656)(5.136875,-0.9147656)(5.1380467,-0.9146875)(5.1404686,-0.9146094)(5.1452346,-0.91453123)(5.154844,-0.91429687)(5.1739845,-0.91382813)(5.176328,-0.91382813)(5.178672,-0.91375)(5.183359,-0.91367185)(5.1927342,-0.9134375)(5.2115626,-0.9130469)(5.212734,-0.9130469)(5.2139063,-0.91296875)(5.21625,-0.91296875)(5.2209377,-0.9128906)(5.2303123,-0.91265625)(5.2491407,-0.91234374)(5.2503905,-0.91234374)(5.251641,-0.9122656)(5.2542186,-0.9122656)(5.269453,-0.91203123)(5.2701564,-0.91203123)(5.270781,-0.9119531)(5.2746096,-0.9119531)(5.2898436,-0.9117188)(5.291172,-0.9117188)(5.292422,-0.91164064)(5.294922,-0.91164064)(5.300078,-0.9115625)(5.3102345,-0.9114063)(5.330625,-0.91117185)(5.3353906,-0.91117185)(5.340156,-0.9110938)(5.3496094,-0.9110156)(5.352031,-0.9110156)(5.354375,-0.9109375)(5.3591404,-0.9109375)(5.368672,-0.9108594)(5.3710155,-0.9108594)(5.3734374,-0.91078126)(5.380547,-0.91078126)(5.3828907,-0.9107031)(5.3924217,-0.9107031)(5.3971877,-0.910625)(5.4125,-0.910625)(5.413125,-0.9105469)(5.443047,-0.9105469)(5.4433594,-0.91046876)(5.4697657,-0.91046876)(5.4710937,-0.9105469)(5.4998436,-0.9105469)(5.5004687,-0.910625)(5.514375,-0.910625)(5.519453,-0.9107031)(5.529609,-0.91078126)(5.5396876,-0.91078126)(5.549844,-0.9109375)(5.554844,-0.9109375)(5.5599217,-0.9110156)(5.570078,-0.9110938)(5.572422,-0.9110938)(5.5747657,-0.91117185)(5.579453,-0.91117185)(5.5889063,-0.91132814)(5.59125,-0.91132814)(5.593672,-0.9114063)(5.5983596,-0.9114063)(5.6078124,-0.9115625)(5.6089845,-0.9115625)(5.610156,-0.91164064)(5.6125,-0.91164064)(5.6171875,-0.9117188)(5.645547,-0.9121875)(5.6467967,-0.9121875)(5.648125,-0.9122656)(5.650625,-0.9122656)(5.6557813,-0.9124219)(5.6660156,-0.9125781)(5.6864843,-0.9130469)(5.6877346,-0.9130469)(5.6889844,-0.913125)(5.6915627,-0.913125)(5.6967187,-0.91328126)(5.706953,-0.9135156)(5.7274218,-0.9140625)(5.728594,-0.9140625)(5.7297654,-0.91414064)(5.7321873,-0.9142188)(5.7369533,-0.91429687)(5.7464843,-0.9146094)(5.765625,-0.91515625)(5.7657814,-0.91515625)(5.7658596,-0.9152344)(5.7679687,-0.9152344)(5.7703905,-0.9153125)(5.7846875,-0.91578126)(5.8038282,-0.91648436)(5.804375,-0.91648436)(5.805,-0.9165625)(5.806172,-0.9165625)(5.8132033,-0.91679686)(5.8225,-0.9171875)(5.84125,-0.9178906)(5.87875,-0.9195313)(5.879375,-0.9195313)(5.88,-0.91960937)(5.88125,-0.91960937)(5.883828,-0.9197656)(5.8990626,-0.92046875)(5.919375,-0.92148435)(5.96,-0.92359376)(5.9603124,-0.92359376)(5.960625,-0.9236719)(5.961172,-0.9236719)(5.9623437,-0.92375)(5.9647655,-0.9238281)(5.9694533,-0.92414063)(5.9789844,-0.9246875)(5.9978905,-0.92578125)(6.0358596,-0.9282031)(6.1180468,-0.9340625)(6.194844,-0.94039065)(6.270078,-0.9475)(6.351719,-0.9561719)(6.427969,-0.9653125)(6.5105467,-0.97632813)(6.5915623,-0.98828125)(6.6671877,-1.0005469)(6.749219,-1.0151563)(6.8257813,-1.0299219)(6.900781,-1.045625)(6.9821873,-1.0640625)(7.058203,-1.0825781)(7.140547,-1.1040626)(7.2214065,-1.1267968)(7.296797,-1.1494532)(7.378594,-1.1755469)(7.4549217,-1.2014843)(7.5297656,-1.2284375)(7.6109376,-1.259375)(7.686719,-1.2899219)(7.768828,-1.3248438)(7.8455467,-1.3592969)(7.920703,-1.3946875)(8.002266,-1.435)(8.07836,-1.4745313)(8.160859,-1.5194532)(8.2417965,-1.565625)(8.317344,-1.610625)(8.317422,-1.610625)(8.317735,-1.6107813)(8.318047,-1.6110157)(8.318672,-1.6114062)(8.32,-1.6121875)(8.322657,-1.6138281)(8.33836,-1.6234375)(8.359375,-1.6364063)(8.359453,-1.6364844)(8.359766,-1.6366407)(8.360078,-1.636875)(8.360703,-1.6372656)(8.362031,-1.6380469)(8.364688,-1.6396875)(8.369922,-1.6429688)(8.38039,-1.6496093)(8.380468,-1.6496093)(8.380781,-1.6497656)(8.381094,-1.65)(8.381719,-1.6503906)(8.383047,-1.65125)(8.385703,-1.6528906)(8.390938,-1.6561719)(8.391094,-1.6563281)(8.39125,-1.6564063)(8.391562,-1.6566406)(8.392265,-1.6570313)(8.393594,-1.6578907)(8.396172,-1.6595312)(8.39625,-1.6596094)(8.396328,-1.6596094)(8.396484,-1.6597656)(8.396875,-1.6599219)(8.3975,-1.6603906)(8.3988285,-1.6611719)(8.398984,-1.6613281)(8.39914,-1.6614063)(8.399453,-1.6616406)(8.400156,-1.6620313)(8.400234,-1.6621094)(8.400312,-1.6621094)(8.400469,-1.6622657)(8.400782,-1.6624218)(8.400937,-1.6625781)(8.4010935,-1.6626563)(8.401172,-1.6627344)(8.40125,-1.6627344)(8.401406,-1.6628907)
\psline[linewidth=0.0139999995cm](0.60546875,-1.0865625)(8.401406,-1.0865625)
\psline[linewidth=0.028800001cm](8.401406,-2.0501564)(0.60546875,-2.0501564)
\psline[linewidth=0.028800001cm](0.60546875,-2.0501564)(0.60546875,2.7679687)
\psline[linewidth=0.028800001cm](0.60546875,2.7679687)(8.401406,2.7679687)
\psline[linewidth=0.028800001cm](8.401406,2.7679687)(8.401406,-2.0501564)
\psline[linewidth=0.028800001cm](0.60546875,-2.0501564)(0.60546875,-1.9701563)
\psline[linewidth=0.028800001cm](2.1646094,-2.0501564)(2.1646094,-1.9701563)
\psline[linewidth=0.028800001cm](3.723828,-2.0501564)(3.723828,-1.9701563)
\psline[linewidth=0.028800001cm](5.2830467,-2.0501564)(5.2830467,-1.9701563)
\psline[linewidth=0.028800001cm](6.8422656,-2.0501564)(6.8422656,-1.9701563)
\psline[linewidth=0.028800001cm](8.401406,-2.0501564)(8.401406,-1.9701563)
\psline[linewidth=0.028800001cm](0.60546875,-2.0501564)(0.68546873,-2.0501564)
\psline[linewidth=0.028800001cm](0.60546875,-1.0865625)(0.68546873,-1.0865625)
\psline[linewidth=0.028800001cm](0.60546875,-0.12289062)(0.68546873,-0.12289062)
\psline[linewidth=0.028800001cm](0.60546875,0.8407031)(0.68546873,0.8407031)
\psline[linewidth=0.028800001cm](0.60546875,1.804375)(0.68546873,1.804375)
\psline[linewidth=0.028800001cm](0.60546875,2.7679687)(0.68546873,2.7679687)
\psline[linewidth=0.028800001cm](0.60546875,2.7679687)(0.60546875,2.6879687)
\psline[linewidth=0.028800001cm](2.1646094,2.7679687)(2.1646094,2.6879687)
\psline[linewidth=0.028800001cm](3.723828,2.7679687)(3.723828,2.6879687)
\psline[linewidth=0.028800001cm](5.2830467,2.7679687)(5.2830467,2.6879687)
\psline[linewidth=0.028800001cm](6.8422656,2.7679687)(6.8422656,2.6879687)
\psline[linewidth=0.028800001cm](8.401406,2.7679687)(8.401406,2.6879687)
\psline[linewidth=0.028800001cm](8.401406,-2.0501564)(8.321406,-2.0501564)
\psline[linewidth=0.028800001cm](8.401406,-1.0865625)(8.321406,-1.0865625)
\psline[linewidth=0.028800001cm](8.401406,-0.12289062)(8.321406,-0.12289062)
\psline[linewidth=0.028800001cm](8.401406,0.8407031)(8.321406,0.8407031)
\psline[linewidth=0.028800001cm](8.401406,1.804375)(8.321406,1.804375)
\psline[linewidth=0.028800001cm](8.401406,2.7679687)(8.321406,2.7679687)
\rput(0.19609372,-0.9728125){1.0}
\rput(0.1848438,-0.0128125){1.1}
\rput(0.19390625,0.9471875){1.2}
\rput(0.19890624,1.9071873){1.3}
\rput(0.19640625,2.8871875){1.4}
\rput(0.20265625,-1.9328126){0.9}
\rput(0.58515626,-2.2928126){0}
\rput(2.1539063,-2.2928126){0.1}
\rput(3.7229686,-2.2928126){0.2}
\rput(5.267969,-2.2928126){0.3}
\rput(6.8254685,-2.2928126){0.4}
\rput(8.394219,-2.2928126){0.5}
\rput(4.433906,-2.8328125){$\bar{L}$}
\psline[linewidth=0.04cm,linecolor=color147](1.0342188,2.3801563)(2.0342188,2.3801563)
\psline[linewidth=0.04cm,linecolor=color148](1.0342188,1.8801563)(2.0342188,1.8801563)
\psline[linewidth=0.04cm,linecolor=color1687](1.0342188,1.3801563)(2.0342188,1.3801563)
\rput(2.790625,2.3701563){3 orders}
\rput(2.7951562,1.8701563){4 orders}
\rput(2.789375,1.3701563){5 orders}
\rput{-270.0}(-0.58125,0.25){\rput(0.1490625,0.0196875){$\sigma^\mathrm{NG}/\exp(\sigma_2^\mathrm{NG})$}}
\end{pspicture}}
\caption{\label{fig2}Plot of the ratio $\sigma^{\mathrm{NG}}/\exp(\sigma_2^\mathrm{NG})$.}
\end{center}
\end{figure} 
the ratio $\sigma^{\mathrm{NG}}/\exp(\sigma_2^\mathrm{NG})$, with $\sigma_2^\mathrm{NG} = -\bar{L}^2/2!\times \mathrm{C_FC_A}\zeta_2$, being the second-order NGLs function in the exponent. In figure \ref{fig2} we show truncations of the series at third, fourth and fifth orders. It is clear from the curves that the third, fourth and fifth-order terms in the exponent form a significant $\mathcal{O}(30\%)$ contribution (particularly at larger values of $\bar{L}$), meaning that the series converges very slowly. It also means that more terms of the series in the exponent are needed for a phenomenologically reliable estimate of the all-orders behaviour of the distribution.

\section{Comparison to numerical results at large $\mathbf{N_c}$}

In this section we compare the results we obtained for the resummed NGLs, the exponential form \eqref{eq:exponential}, with the numerical all-orders resummed result obtained by Dasgupta and Salam at large $\mathrm{N_c}$ via a Monte Carlo approach \cite{Dasgupta:2001sh}. Their numerical result is parameterised as follows \cite{Dasgupta:2001sh}:
\begin{align}
\sigma^\mathrm{NG}_\mathrm{DS}(t) & = \exp\left(-\mathrm{C_FC_A}\frac{\pi^2}{3}\frac{1+(0.85 \mathrm{C_A}t)^2}{1+(0.86\mathrm{C_A}t)^{1.33}}t^2\right), \label{eq:sigmaDS}
\end{align}
with the evolution parameter $t$ given by:
\begin{align}\label{eq:evolution}
t= & \frac{1}{4\pi\beta_0} \ln \frac{1}{1-2\beta_0\alpha_s L}\,,
\end{align}
and $\beta_0$ is the leading-order coefficient of the QCD $\beta$ function. We show in figure \ref{fig:largenc} (on the left) a plot of the functions $\sigma_\mathrm{DS}^\mathrm{NG}$ (eq. \eqref{eq:sigmaDS}) and $\sigma^\mathrm{NG}$ (eq. \eqref{eq:exponential}) with various truncations of the NGLs series in the exponent of eq. \eqref{eq:exponential}. We also show in the same figure (on the right) a plot of the ratio of the two functions.
\input{Figs/fig3.tex}

It is clear from the plots, particularly the right-hand-side one, that as one adds more terms in the exponent of eq. \eqref{eq:exponential}, one obtains larger intervals (starting from $t=0$ and spanning over large values of $t$) over which there is an agreement between the analytical form \eqref{eq:exponential} and the parameterised form of the Monte Carlo output \eqref{eq:sigmaDS}. This observation means that it may suffice to compute just a few more higher-order terms in order to obtain agreement between the two functions for a phenomenologically sufficient interval of $t$. It is worth noting that the second-order result has the peculiar feature that it represents the best fit to the all-orders result over the full range of $t$ considered.

\section{Summary and outlook}

In this work we addressed the calculation of NGLs at finite $\mathrm{N_c}$ at single log accuracy up to fifth order for hemisphere mass distribution in $e^+e^-\to$ 2 jets. This was achieved through a brute-force method in which we integrated eikonal squared amplitudes over the appropriate phase space. We observed that the obtained series of NGLs suggest a possible resummation into an exponential form. When expanded to leading order in colour, our results exactly reproduce those obtained at large $\mathrm{N_c}$ by means of the analytical solution to the BMS equation. The results we obtained agree with the statement made in ref. \cite{Hatta:2013iba} that finite-$\mathrm{N_c}$ contribution forms a small correction to the large-$\mathrm{N_c}$ result, meaning that the large-$\mathrm{N_c}$ approximation is a good one, at least in the context of $e^+e^-$ collisions.

Our next task in this work is to go beyond fifth order in the calculation of the series of NGLs in an attempt to confirm the structure of the resummation into an exponential form. In addition we plan to extend this work by analytically investigating the effect of jet clustering on the hemisphere mass as well as other jet-shape distributions.

\acknowledgments
This work is supported in part by CNEPRU Research Project D01320130009.

\end{document}